\documentclass[preprint,prb,doublespace,showpacs,preprintnumbers,amsmath,
amssymb]{revtex4}
\usepackage{latexsym}
\usepackage{graphicx}
\usepackage{dcolumn}

\begin{document}

\title{Size distribution of sputtered particles from
Au nanoislands due to MeV self-ion bombardment}

\author{B. Satpati, J. Ghatak, P. V. Satyam and B. N.
Dev\footnote{email: bhupen@iopb.res.in, FAX: +91-674-2300142}}

\address{Institute of Physics, Sachivalaya Marg, Bhubaneswar-751005, India}

\begin{abstract}

Nanoisland gold films, deposited by vacuum evaporation of gold onto
Si(100) substrates, were irradiated with 1.5 MeV Au$^{2+}$ ions up to a fluence
of $5\times 10^{14}$ ions cm$^{-2}$ and at incidence angles up to
$60^{\circ}$ with respect to the surface normal. The sputtered particles
were collected on carbon coated grids (catcher grid) during ion
irradiation and were analyzed with transmission electron microscopy and
Rutherford backscattering spectrometry. The average sputtered particle
size and the areal coverage are determined from transmission electron
microscopy measurements, whereas the amount of gold on the substrate is
found by Rutherford backscattering spectrometry. The size distributions of
larger particles (number of atoms/particle, $n$ $\ge$ 1,000)  show an
inverse power-law with an exponent of $\sim$ -1 in broad agreement with a
molecular dynamics simulation of ion impact on cluster targets.
\end{abstract}

\maketitle

\section{INTRODUCTION} Although emission of clusters of atoms during ion
sputtering was first observed almost half a century ago \cite{Honig}, this
phenomenon has attracted more attention in recent years, both
theoretically and experimentally. Recent reviews on this topic are
available in the literature. \cite{Hofer,Nordlund,Wucher0} Various
scenarios for cluster emission have been projected; these range from
linear-cascade sputtering over surface evaporation to gasification (or
volume evaporation). \cite{Reimann}

For cluster emission, it has been shown \cite{Coon,Wucher1} that the size
distribution, $Y(n)$, of the emitted clusters obey an inverse power-law:

\begin{equation} 
Y(n) \propto n^{-\delta}
\label{power-law} 
\end{equation}

\noindent where $n$ is the number of atoms in a given cluster.  For small
emitted clusters ($n$ $\lesssim$ 40) the reported values of $\delta$ range
between 4 and 8. \cite{Coon,Wucher1} For larger clusters ($n$ $\ge$ 500)
recently Rehn {\it et al.} \cite{Rehn} reported a value of $\delta$ = 2,
which is consistent with the model of Bitensky and Parilis.
\cite{Bitensky} This value of $\delta$ = 2 is consistent with the
mechanism that large clusters are produced when shock waves, generated by
subsurface displacement cascades, ablate the surface. However, these
results are for sputtering either from solid surfaces or from continuous
films. For nanodispersed films one may expect different results as the
sputtering phenomenon can be expected to be considerably stronger in small
finite systems as nanometer sized islands. Kissel and Urbassek
investigated sputtering from clusters by molecular-dynamics (MD) computer
simulation where they modeled the materials effects occurring in a highly
energized nonequilibrium cascade volume (i.e. the effect of energy spike).
\cite{Kissel} In this simulation they used 100 keV Au atom bombardment of
spherical Au clusters (of 4 nm radius). Their results show a distribution
of emitted clusters containing up to 100 atoms and even beyond. Smaller
clusters (up to $ n \sim$10) show an inverse power-law [Eq.
\ref{power-law}] with $\delta$ = 2.3. For larger cluster sizes, the decay
appears to become even softer. This fall-off is distinctly slower than
that observed for keV-ion bombardment of metal surfaces.
\cite{Wahl,Staudt} This means that larger clusters are emitted with a
higher probability from cluster targets compared to targets with planar
solid surfaces. For the larger sizes ($n$ $>$ 10) of emitted clusters, the
value of $\delta$ was not extracted from the simulation results.

To our knowledge, no experimental results of size distribution of emitted
particles in ion bombardment of free clusters are available. However, ion
bombardment of clusters deposited on surfaces has recently been
investigated experimentally. \cite{Baranov,Satpati2}, although no effort
was made to verify the power-law dependence of $Y(n)$ and extract the
value of the exponent. In this paper we present an effort to verify the
power-law dependence of the large ($n \ge$ 1,000) emitted particles from
nanodispersed Au on silicon substrates due to MeV Au ion bombardment.

In earlier experiments, where particle ejection from ion-bombarded
nanoisland films was investigated, cluster ion beams instead of a monomer
ion beam was used. \cite{Baranov} The interaction of a cluster ion beam
with solids may be quite different (nonlinear) in comparison to a monomer
beam; this happens due to vicinage of atoms in the cluster beam.
\cite{Sanjit} Thus the interaction of a cluster ion beam with nanoisland
targets can in principle, produce a different distribution of ejected
particles compared to the interaction of a monomer ion beam. Here we
consider the simpler case of irradiation with a monomer ion beam. Our
present study is concerned with the interaction of 1.5 MeV Au$^{2+}$ (monomer)
ions with Au nanoisland films.

Our earlier studies have shown that when nanodispersed targets are
bombarded by 1.5 MeV Au ions, the size and coverage of sputtered particles
collected on a catcher grid are larger compared to thicker continuous film
targets. \cite{Satpati2} We observed also higher probability of crater
formation in island thin films compared to thick continuous
films due to MeV self-ion irradiation and proposed it as experimental
evidence of the energy spike confinement in nanoisland targets.
\cite{Satyam,Satpati3,Satpati4} An energy spike confinement in the
nanoislands can result in either a thermal spike or a shock wave spatial
confinement.

\section{EXPERIMENTAL} 

Au films of thickness 1.3 $\pm$ 0.1 nm were deposited by thermal
evaporation under high-vacuum conditions ($\approx 2\times10^{-6}$ mbar)
on native-oxide covered Si(100) substrates at room temperature (RT).
Deposition rate was 0.1 nm/s. Before thin film deposition, substrates were
ultrasonically cleaned in acetone, methanol, trichloroethylene, methanol,
deionized water and in acetone sequentially. Ion sputtering experiments
were performed with 1.5 MeV Au$^{2+}$ ions with fluences ranging from
$1\times10^{12}$ to $5\times10^{14}$ ions cm$^{-2}$ at different angles of
incidence with respect to the surface normal. The incident ion current
(over a scanned beam area of 1$\times$1 cm$^2$) was kept between 20 to 40
nA under secondary electron suppressed geometry. During irradiation,
the sputtered particles were collected on a catcher grid that was
positioned $\approx $1 cm above the target, the catcher surface making an
angle ($\approx15 ^{\circ}$) with respect to the sample surface.  Such a
large distance of the catcher is to reduce overlap of the collected
particles on the catcher grid. During irradiation, $\approx$
2$\times$10$^{-7}$ mbar pressure was maintained in the irradiation
chamber. Transmission electron microscopy (TEM) measurements were
performed (using a JEOL JEM-2010 (UHR) electron microscope operating at
200 kV) on the ion-bombarded samples to study the changes in the
morphology of the nanostructures on the substrates and to determine the
surface coverage and the size distribution of the sputtered particles on
the catcher grids. The effective thickness of the as-deposited Au films on
Si substrates has been determined by Rutherford backscattering
spectrometry (RBS) using the bulk atomic density of Au and the RUMP
simulation package.  \cite{Doolittle}

\section{RESULTS AND DISCUSSIONS} 

The deposited Au grows as islands on native-oxide covered Si substrate.
Fig. 1 shows a plan-view TEM micrograph of such a sample and a histogram
for the particle size distribution. The distribution is fitted well by a
Gaussian function. It is found that the Au islands are isolated with an
average particle size of 11.1 $\pm$ 0.1 nm with a standard deviation,
$\sigma$ = 5.1 $\pm$ 0.3 nm and the surface coverage of islands is
$\sim$26\%. These values are obtained using several TEM micrographs and
the ImageJ analysis package. \cite{ij}

In the description of the following results, the as-deposited nanoisland
Au films that have been ion-bombarded will be called targets. Figs. 2(a) and
(b) show TEM images from the target and from the catcher grid, respectively,
after irradiation (fluence: $1 \times 10^{14}$ ions cm$^{-2}$) 
at 0$^{\circ}$-impact angle (with respect to the surface normal).
Fig. 2(c) is a TEM image from the same region as in 2(b) of the catcher grid 
obtained by some
underfocusing of the objective lens. It is evident from Fig. 2(c) 
that the sputtered Au particles are three-dimensional (spherical in most 
cases) in nature. The particle size distribution on
the catcher grid is shown in Fig. 2(d). The frequency distribution of the
particle size in Fig. 2(d) and in the latter figures are obtained by
measuring particle size from many TEM images ($\sim$ 700 particles) from 
each catcher grid.
Following ion irradiation, the surface coverage of islands on the target
decreases from 26\% to 7\%.

Figs. 3(a), (b), (c), and (d) show plan-view TEM micrographs from the
target after irradiation at 30$^{\circ}$-impact angle with fluences 
of $1\times10^{12}$, $1\times10^{13}$,
$1\times 10^{14}$ and $5\times10^{14}$ ions cm$^{-2}$, respectively.  It
appears from the micrographs that in the initial stage of irradiation,
particles get agglomerated and elongated. For an ion fluence of
$1\times10^{12}$ ions cm$^{-2}$, surface coverage of islands increases
from 26\% to 34\% and then it reduces to 18\% for a fluence of
$1\times10^{13}$ ions cm$^{-2}$ and 9\% for a fluence of $1\times10^{14}$
ions cm$^{-2}$. Finally, at a fluence of $5\times10^{14}$ ions cm$^{-2}$,
islands are embedded into the Si substrate.

Figs. 3(e), (f), (g), and (h) show plan-view TEM micrographs of sputtered
particles collected on catcher grids as a result of irradiating the target
to the fluences of $1\times10^{12}$, $1\times10^{13}$, $1\times 10^{14}$
and $5\times10^{14}$ cm$^{-2}$, respectively. The corresponding surface
coverage of islands are 2.6\%, 5.1\%, 6.2\% and 19.5\%, respectively.  
They correspond to the targets in (a), (b), (c), and (d), respectively.  
The corresponding particle size distributions on the catcher grids are
shown in (i), (j), (k), and (l), respectively. The size distribution can be
fitted to a log-normal function as described later. The embedding process 
already starts at a fluence of $1\times
10^{14}$ ios cm$^{-2}$. \cite{Satpati1} The slightly smaller particle size
observed for the highest fluence (see Table-1), is perhaps an indication
that the probability of emission of smaller particles is higher from the
embedded particles on the target.

Figs. 4(a), (b), and (c) show plan-view TEM micrographs from the target
after irradiation at 60$^{\circ}$-impact angle with fluences of
$1\times10^{12}$, $1\times10^{13}$ and $1\times 10^{14}$ ions cm$^{-2}$,
respectively. For the fluence of $1\times10^{12}$ ions cm$^{-2}$, the
surface coverage of islands increases from 26\% (for as-deposited sample)
to 37\% and then reduces to 12\% for $1\times10^{13}$ ions cm$^{-2}$, and
finally at a fluence of $1\times10^{14}$ ions cm$^{-2}$, islands are
embedded into the Si substrate. Compared to 30$^{\circ}$-impact angle,
here the surface coverage of Au particles on the target reduces more
rapidly with ion fluence. The embedding also begins to occur at a smaller
fluence for the 60$^{\circ}$-impact angle of the ion beam. It appears that
embedding has already begun at a fluence of $1\times10^{13}$ ions
cm$^{-2}$. Figs. 4(d), (e), and (f) show plan-view TEM micrographs of
sputtered particles collected on catcher grid. They correspond to the
targets in (a), (b), and (c), respectively. The particle size
distributions obtained from the catchers are shown in (g), (h), and (i),
respectively.

From the cross-sectional TEM images in Fig. 5 it is evident that during
ion irradiation the nanoparticles initially (for fluence:
1$\times$10$^{12}$ ions cm$^{-2}$) spread over the surface. At a higher
fluence (1$\times$10$^{13}$ ions cm$^{-2}$) agglomeration is dominant and
larger islands are produced. At a fluence of 1$\times$10$^{14}$ ions
cm$^{-2}$ almost all the islands get embedded into the Si substrate.
All the embedded particles [Fig. 5(d)] appear to be large. However, one
small particle is seen in Fig. 5(d), which is not embedded. It is also
evident that at lower fluences when the particle size is small, there
is no embedding. This raises a question: is there any critical size of
islands on the target so that there is no embedding below this size ?
This aspect is to be addressed in future investigations.

The size distribution of islands on the as-deposited sample [Fig. 1(b)] is
fitted well by a Gaussian function. The size distributions of islands,
collected on the catcher grid following ion irradiation of the
as-deposited sample, [Figs. 2(d), 3(i)-(l) and 4(g)-(i)] are found to fit
a log-normal function :

\begin{equation}
P(x)=\frac{1}{\sqrt{2\pi}xw} e^{ -\left(\frac{ln(x/x_c)}{\sqrt{2}w}\right)^2}
\end{equation}

\noindent where, $x_c$ is the most probable size and $w$ is
the width of the size distribution of the particles. By fitting
the frequency plot using Eq. 2, we have estimated the most probable
particle size $x_c$ and width $w$. The results are presented in Table 1. 
In order to confirm that the collected
nanoparticles on the catcher grid are not Si clusters, we irradiated a
bare Si substrate (without any deposited Au)  to a fluence of
$1\times10^{14}$ ions cm$^{-2}$. However, no sputtered particle was
observed on the catcher grid in this case. In addition, we have carried
out lattice imaging by high-resolution TEM to confirm that the sputtered
particles are Au nanoclusters.

The sputtered nanoparticles on the catcher grid are usually spherical (as
described earlier). For the lowest ion fluence ($1\times10^{12}$ ions
cm$^{-2}$), the density of islands on the catcher grid [Figs. 3(e) and
4(d)] is low and the island size distribution, although fitted by a
log-normal function, actually does not deviate much from a Gaussian
function. For higher ion fluences, apparently there is ejection of more
larger sized particles [cf. for example, Figs. 3(f), (j) and 4(e), (h)].
This appears to be correlated with the increase of particle size on the
target [cf. Figs. 3(b) and 4(b)]. This correlation has earlier been
observed by Baranov {\it et al.} \cite{Baranov}, who performed experiments
on nanodispersed targets with selected particle sizes.  Thus, for ion
bombardment at higher fluences the size distributions develop a tail
extending into larger sizes and a log-normal function fits the
distribution better.

In order to understand the mechanism involved in the ejection of particles
in ion-target interaction, many authors have predicted different
power-laws. According to Eq. \ref{power-law}, the size distribution,
$Y(n)$, of the sputtered particles obeys an inverse power-law.
\cite{Coon,Wucher1} The value of the exponent $\delta$, in Eq.
\ref{power-law} depends on the mechanism of ion-target interaction.
Experimental determination of the value of $\delta$ exists in the
literature only for solid surfaces and continuous film targets. There are
conflicting results in the literature. The exponent, $\delta$, was found
to correlate \cite{Wucher0,Coon,Wucher1} with the total sputtering yield,
such that higher sputtering yields result in smaller values of $\delta$.
Whereas Rehn {\it et al.} \cite{Rehn} reported that the value of $\delta$
is independent of sputtering yield. Using mass-spectrometry techniques,
the observed values of $\delta$ range between 4 and 8. \cite{Coon,Wucher1}
However, these values are for smaller ejected clusters of $n \lesssim$
40 atoms. A high exponent means that smaller clusters are present with
high probabilities in the mass-spectrometry studies. It is suspected that
some large clusters get fragmented to increase the number of small
clusters. Rehn {\it et al.} \cite{Rehn} restricted their studies to large
clusters ($n \ge$ 500) and obtained a value of $\delta$ = 2. Consistent
with a theoretical model this value of the exponent indicates that the
large clusters are produced when shock waves, generated by subsurface
displacement cascades, ablate the surface.  The shock wave model proposed
by Bitensky and Parilis \cite{Bitensky} predicts the inverse-square
dependence of $Y(n)$. On the other hand no simple theoretical model has
been able to reproduce both the power-law and high values of $\delta$
found from mass-spectrometry techniques. All these results discussed in
this paragraph are for solid surfaces and continuous thin film targets.

For cluster targets MD simulation of Kissel and Urbassek \cite{Kissel}
predicts a value of $\delta$ $\cong$ 2.3 for small clusters ($n \lesssim$
10). For larger clusters ($n >$ 10) the decay appears to become softer,
although they have not tried to extract the value of $\delta$ from this
region. To our knowledge, this is the only theoretical work where cluster
ejection in an ion-cluster interaction has been considered.  We try to
compare our results with these simulation results. We have fitted the
larger cluster region ($n >$ 10) of the spectrum from Ref. 10 and obtained
a value of $\delta \approx$ 1.15. The plot of data with the fit is shown
in Fig. 6.

Now we present our results of size distributions of the larger clusters
($n \ge$ 1,000) in terms of the inverse power-law dependence of $Y(n)$ on
$n$. The particle size distribution, $Y(n)$, i.e. the number of clusters
containing $n$ atoms, was determined from the catcher grid as follows. The
size of each particle was determined using ImageJ software. The measured
lateral dimensions were converted into the number of atoms for a given
cluster by assuming that the particles are spherical with the density of
bulk Au. All visible particles in an area of $\approx 3\times 10^5$ nm$^2$  
(ten frames, each of $\sim$ 210 $\times$ 160 nm$^2$) were sized. 
The particles were grouped into bins having 1 nm steps.
The procedure was repeated for each case of irradiation. The size
distributions that were obtained in this manner are displayed in Fig. 7
alongwith a power-law fit. The values of the exponent $\delta$, although
show some variations, broadly are close to that of the theoretical value
shown in Fig. 6.  Nevertheless, one notices that for the lowest fluence
($1\times10^{12}$ ions cm$^{-2}$) where the size distributions of the
emitted particles [Fig. 3(i) and 4(g)] are nearly Gaussian and the
presence of larger size particles is not prominent, the value of the
exponent $\delta$ is larger. One may also note that the particle sizes on
the target in these cases are also smaller.

Studies of ejected particle distribution from nanoisland targets must be
explored further both theoretically and experimentally. In the simulation
of Kissel and Urbassek \cite{Kissel}, electronic stopping of the ions in
the target island has been disregarded. The electronic energy loss ($S_e$)
and the nuclear energy loss ($S_n$) for 1.5 MeV Au ions in a Au target are
2.41 keV/nm and 9.53 keV/nm respectively, as obtained by using the SRIM
2003 code. \cite{srim} It is important to note that, since $S_n$ is
$\approx$ 3.9 $S_e$, although the effect of $S_n$ will be dominant,
contributions to sputtering from both mechanisms could be possible. Some
MD simulations show a dependence of sputtering yield on $S_e$.
\cite{Bringa} It is to be investigated how the inclusion of the effects of
both $S_n$ and $S_e$ affect the value of $\delta$.

In order to explain the nature of the observed size distributions (Fig.
7), i.e., $Y(n)$ rising at the beginning, following inverse power-law at
the middle and then falling faster at the end. We refer to the work of
Baranov {\it et al.}. \cite{Baranov} They have reported sputtering of
nanodispersed targets and shown that the size distributions of ejected
clusters are systematically shifted towards smaller sizes in comparison
with the island size distributions on the targets.  For example average
island sizes of 4.5$\pm$1.2 nm and 6.7$\pm$2.5 nm on the target led to
average island sizes of 3.6$\pm$1.1 nm and 4.4$\pm$1.6 nm, respectively,
of the ejected particles. So, experimentally the general trend is that
smaller particles on the target produces smaller ejected particles and the
larger particles on the target produces larger ejected particles. The
simulation of Kissel and Urbassek \cite{Kissel} shows that for a target
particle of a given size, a size distribution of ejected particles is
generated. Combining these two results (experimental and theoretical) one
would expect that when the target particle is larger a distribution of
particles with the same exponent but shifted to a larger average size
would be generated. Using this concept we can explain the nature of our
observed $Y(n)$ versus $n$ curves. We have illustrated this in Fig. 8. We
have generated several plots with the same exponent and then shifted the
curves along the $n$-axis to mimic contributions from particles of
different sizes, D (D$_1$ $<$ D$_2$ $<$ D$_3$ $<$ D$_4$ $<$ D$_5$), 
representing the diameter of the particle on the target. Then we 
added all the sets of data. The
resultant curve (solid curve) shows the same nature as our experimental
curves. This explains why with increasing $n$ experimental data show the
rising nature of $Y(n)$ at the beginning, power-law fall in the middle and
a faster fall at the end. This behavior is prominently seen in Fig. 7(a)  
[$1\times10^{14}$ ions cm$^{-2}$] and in Fig. 7(b)  [$1\times10^{13}$ ions
cm$^{-2}$]. Kissel and Uebassek have done the simulation only
for a single size particle (4 nm radius). It would be useful to have
simulation results for different sizes of target islands.

\section{CONCLUSIONS}

We have investigated the distribution of ejected (sputtered) particles from
nanoislands in monomer ion interaction with nanodispersed solid targets.
The ejected cluster size distribution as a function of number of atomsper ejected particle shows a
power-law decay,  $Y(n) \sim n^{-\delta}$, with $\delta \approx 1$. The
exponent is in broad agreement with a molecular dynamics simulation where
the effect of an energy spike in the target islands is considered. In the
simulation, as neither serious effort has been made to extract the exponent
for larger ejected islands nor the size dependence of target islands has
been considered, more theoretical input is required to understand the
distribution of sputtered particles and the mechanism of sputtering.

\section{ACKNOWLEDGMENTS} The authors would like to thank Dr. T. Som for
help in RBS measurements and all the accelerator staff at the Ion Beam
Laboratory, Institute of Physics, Bhubaneswar.

\newpage
FIGURE CAPTIONS

\vspace*{0.2in} \noindent FIG. 1:(a) A plan-view TEM image from an
as-deposited film (effective thickness: 1.3 nm) showing Au nanoparticles
on a Si substrate and (b) histogram showing particle size distribution.  
The histogram shown in (b) is fitted with a Gaussian distribution to
determine the average particle size ($x_c$) and the standard deviation
($\sigma$) in the size distribution.

\vspace*{0.15in} \noindent FIG. 2: Plan-view TEM images:  (a) Au particles
on the target following 1.5 MeV Au$^{2+}$ (fluence: $1\times10^{14}$ ions
cm$^{-2}$) bombardment at an impact angle of 0$^{\circ}$, (b) sputtered
particles on the catcher grid, (c) sputtered particles on catcher grid
showing spherical nature of the particles after some underfocousing of the
objective lens. (d) histogram showing size distribution of sputtered
particles.

\vspace*{0.15in} \noindent FIG. 3: Plan-view TEM images from the target
(a, b, c, d) following 1.5 MeV Au$^{2+}$ bombardment with an impact angle
of 30$^{\circ}$, from the catcher grid (e, f, g, h) and histograms (i, j,
k, l) showing the size distribution of the sputtered particles on the
corresponding catcher grids. Each row in the figure corresponds to a given
ion fluence:  (a), (e), (i) - $1\times10^{12}$ ions cm$^{-2}$;  (b), (f),
(j) - $1\times10^{13}$ ions cm$^{-2}$;  (c), (g), (k) - $1\times10^{14}$
ions cm$^{-2}$;  and (d), (h), (l) - $5\times10^{14}$ ions cm$^{-2}$.

\vspace*{0.15in} \noindent FIG. 4: Plan-view TEM images from the target
(a, b, c) following 1.5 MeV Au$^{2+}$ bombardment at an impact angle of
60$^{\circ}$, from the catcher grid (d, e, f) and histograms (g, h, i)
showing the size distribution of the sputtered particles on the
corresponding catcher grids. Each row in the figure corresponds to a given
ion fluence:  (a), (d), (g) - $1\times10^{12}$ ions cm$^{-2}$;  (b), (e),
(h) - $1\times10^{13}$ ions cm$^{-2}$;  and (c), (f), (i) -
$1\times10^{14}$ ions cm$^{-2}$.

\vspace*{0.15in} \noindent FIG. 5: XTEM images of from the target for the
case of 60$^{\circ}$- impact angle of ion bombardment: (a) an as-deposited
nanodispersed Au film; (b), (c) and (d) from films irradiated with a
fluence of 1$\times$10$^{12}$ ions cm$^{-2}$, 1$\times$10$^{13}$ ions
cm$^{-2}$ and 1$\times$10$^{14}$ ions cm$^{-2}$, respectively.

\vspace*{0.15in} \noindent FIG. 6: Data taken from Ref.10 and fitted with
a power-law given in Eq. 1, to get the exponent ($\delta$)  for larger
sputtered particles.

\hspace*{2.0in}

\vspace*{0.15in} \noindent FIG. 7: Measured number of collected
nanoparticles as a function of nanoparticle size (n) for 1.5 MeV Au ion
irradiation: (a) for its 30$^{\circ}$-impact with various fluences shown
in Fig. 3(e)-(h); (b) for its 60$^{\circ}$-impact with various fluences
shown in Fig. 4(d)-(f); (c) for a fluence of $1\times10^{14}$ ions
cm$^{-2}$ for various impact angles.

\vspace*{0.15in} \noindent FIG. 8: Illustration of the nature of the
observed distribution of the ejected particles [$Y(n)$ vs. $n$ curves]:  
Plots of the power-law-generated data for five different particle sizes
(diameter) on
the target [D$_1$ $<$ D$_2$ $<$ D$_3$ $<$ D$_4$ $<$ D$_5$]. These five
sets of data were added to obtain the resultant yield to explain the
nature of our experimental data, i.e. rising nature at the beginning, a
power-law decay in the middle and a faster fall at the end (see text for
details).

\newpage
\begin{table}
\caption{Size distribution of the sputtered particles for different
fluences and impact angles}
\begin{center}
\begin{tabular}{|c|c|c|}
\hline
Fluence, $\phi$ & Average Size (30$^{\circ}$) & Average
Size (60$^{\circ}$) \\
(ions cm$^{-2}$)& {nm} & {nm}\\
\hline
$1\times10^{12}$ & $x_c$: 7.61$\pm$0.86 $w$: 0.22$\pm$0.17 & $x_c$:
6.28$\pm$0.64 $w$: 0.30$\pm$0.13 \\

$1\times10^{13}$ & $x_c$: 7.67$\pm$0.80 $w$: 0.33$\pm$0.14 & $x_c$:
7.00$\pm$1.39 $w$: 0.60$\pm$0.44 \\

$1\times10^{14}$ & $x_c$: 7.54$\pm$0.85 $w$: 0.35$\pm$0.15 & $x_c$:
6.24$\pm$1.06 $w$: 0.51$\pm$0.28 \\

$5\times10^{14}$ & $x_c$: 7.07$\pm$0.85 $w$: 0.45$\pm$0.14 & ------ \\
\hline
\end{tabular}
\end{center}
\end{table}

\end{document}